\documentclass[twocolumn,superscriptaddress,aps,prl,show pacs]{revtex4-1}
\usepackage{graphicx}
\usepackage{dcolumn}
\usepackage{amsmath}
\usepackage{amssymb}
\usepackage{amsfonts}
\usepackage{bm}
\usepackage{color}
\newcommand{\be}{\begin{equation}}
\newcommand{\ee}{\end{equation}}
\newcommand{\ba}{\begin{eqnarray}}
\newcommand{\ea}{\end{eqnarray}}

\begin{document}
\title{Phase behavior near and beyond\\the thermodynamic stability threshold}
\author{Gianpietro Malescio}
\email[Email: ]{malescio@unime.it}
\affiliation{Universit\`a degli Studi di Messina, Dipartimento di Fisica e di Scienze della Terra, Contrada Papardo, I-98166 Messina, Italy}
\author{Santi Prestipino}
\email[Corresponding author. Email: ]{sprestipino@unime.it}
\affiliation{Universit\`a degli Studi di Messina, Dipartimento di Fisica e di Scienze della Terra, Contrada Papardo, I-98166 Messina, Italy}
\affiliation{CNR-IPCF, Viale F. Stagno d'Alcontres 37, I-98158 Messina, Italy}

\date{\today}

\begin{abstract}
The phase behavior of stabilized dispersions of macromolecules is most easily described in terms of the effective interaction between the centers of mass of solute particles. For molecules like polymer chains, dendrimers, etc., the effective pair potential is {\em finite} at the origin, allowing ``particles'' to freely interpenetrate each other. Using a double-Gaussian model (DGM) for demonstration, we studied the behavior of the system as a function of the attraction strength $\eta$. Above a critical strength $\eta_{\rm c}$, the infinite-size system is Ruelle-unstable, in that it collapses to a cluster of finite volume. As $\eta_{\rm c}$ is approached from below, the liquid-vapor region exhibits an anomalous widening at low temperature, and the liquid density apparently diverges at the stability threshold. Above $\eta_{\rm c}$, the thermodynamic plane is divided in two regions, differing in the value of the {\em average} waiting time for collapse, being finite and small on one side of the boundary line, while large or even infinite in the other region. Upon adding a small hard core to the DGM potential, stability is fully recovered and the boundary line is converted to the spinodal line of a transition between fluid phases. We argue that the destabilization of a colloidal dispersion, as induced by the addition of salt or other flocculant, finds a suggestive analogy in the process by which a strengthening of the attraction pushes a stabilized-DGM system inside the fluid-fluid spinodal region.
\end{abstract}

%   64.75.Xc: Phase separation and segregation in colloidal systems
%   82.70.Dd: Colloids
%   64.60.My: Metastable phases
%   64.70.-p: Specific phase transitions
\maketitle

A major problem in the preparation of a colloid is stabilization, {\it i.e.}, ensuring a homogeneous dispersion of colloidal particles in the solvent. The widespread mechanism by which destabilization of a colloidal dispersion occurs is through the formation of particle clusters of increasing sizes. Of a similar kind is the phenomenon of amyloid aggregation underlying many forms of age-related brain diseases in man~\cite{Pulawski}. Irreversible aggregation can be triggered by changes in the composition of the dispersion, {\it e.g.}, by adding salt or depletants to the colloid. While in general the existence of an impenetrable core prevents colloidal particles from collapsing, in systems like polymer chains, dendrimers, polyelectrolytes, etc., the effective two-body repulsion, resulting from integrating out the internal degrees of freedom of each molecule, is {\em finite} at the origin, implying that the centers of mass of two macromolecules may even coincide~\cite{Louis,Likos1,Likos2,Likos3}. It follows that, except possibly for the late stages of the aggregation process, the destabilization of a dispersion of such particles is analog, at a schematic level, to the destabilization of a bounded pair potential when its attractive component, arising from dispersion forces or depletion mechanisms, becomes sufficiently strong.

As early as fifty years ago, Ruelle and Fisher~\cite{Ruelle,Fisher} observed that a pair potential which is bounded at the origin and enough attractive for some distances undergoes a thermodynamic catastrophe, {\it i.e.}, particles collapse to a finite volume of space. Ruelle and Fisher also provided a simple recipe for calculating the stability threshold. However, the exact terms in which the transition occurs from the thermodynamically stable to the thermodynamically unstable regime as well as the peculiarities of system behavior beyond the threshold of thermodynamic stability were left unresolved. In Refs.\,\cite{Fantoni1,Fantoni2} it was suggested that the unstable regime might actually be composed of two regions, a strongly-unstable region where the phase of the system is non-extensive, and a weakly-unstable region where the system is, to all practical purposes, stable.

To be specific, we inquired into the behavior of particles interacting through a Gaussian pair repulsion augmented with a weaker Gaussian attraction at larger distances~\cite{Prestipino1,Speranza}:
\be
u(r)=\epsilon\exp\left[-\left(\frac{r}{\sigma}\right)^2\right]-\eta\exp\left[-\left(\frac{r}{\sigma}-3\right)^2\right]\,,
\label{eq1}
\ee
where $\epsilon>0$ and $\sigma$ are arbitrary energy and length units, and $\eta>0$. We refer to this model as the {\em double-Gaussian model} (DGM). According to Ref.~\cite{Ruelle}, a sufficient condition for thermodynamic instability is $\widetilde{u}(0)<0$, $\widetilde{u}(k)$ being the Fourier transform of $u(r)$. Conversely, if $\widetilde{u}(k)\ge 0$ for all $k$, then the system is stable. For the DGM potential, we find $\widetilde{u}(0)<0$ for $\eta>\eta_{\rm c}\equiv 0.0263157908\ldots$ (in $\epsilon$ units); below $\eta_{\rm c}$, $\widetilde{u}(k)$ is positive definite. Hence the DGM system is stable for $\eta<\eta_{\rm c}$ and unstable for $\eta>\eta_{\rm c}$. In order to analyze the DGM behavior as a function of the attraction strength $\eta$, all the way to the stability threshold and beyond, we employed a number of simulation techniques, namely Gibbs ensemble Monte Carlo (GEMC), $NVT$ Metropolis Monte Carlo (MC), and molecular dynamics (MD)~\cite{Frenkel}, as well as the hypernetted-chain (HNC) equation of liquid-state theory~\cite{Hansen}. Clearly, for a thermodynamically unstable system it should be carefully checked whether a ``temporal'' window exists or not in which MC averages of macroscopic variables take the same values as MD averages. The important quantity to look at is the average system energy: a total energy proportional to the particle number $N$ is the hallmark of (meta)stable equilibrium, whereas an energy value scaling faster than $N$, typically as $N^2$, is the imprint of system collapse.

For a quick survey of DGM behavior we resorted to the HNC equation, which is quite effective in describing the thermodynamics and structure of particles interacting through a bounded pair potential~\cite{Likos3}. In particular, we looked at the boundary line (BL) separating the region of thermodynamic parameters where the HNC equation can be solved (``stable'' region) from the (``unstable'') region where no solution is found. Upon approaching the BL from the stable region, the computed isothermal compressibility $K_T$ tends to a large positive value. For ordinary simple fluids, characterized by an unbounded short-range repulsion, the BL roughly corresponds to the liquid-vapor spinodal line, marking the threshold of instability towards phase separation. Recently, the BL was computed for the penetrable-square-well model~\cite{Malescio}. When compared to numerical simulations~\cite{Fantoni1,Fantoni2}, the HNC equation yielded fairly reasonable results. 

In distribution-function theories, $K_T$ follows from the relation:
\be
\rho k_BTK_T=1+\rho\widetilde{h}(0)=\frac{1}{1-\rho\widetilde{c}(0)}\,,
\label{eq2}
\ee
where $\rho$ is the number density, $T$ is the temperature, $k_B$ is Boltzmann's constant, $h(r)$ is the total correlation function, and $c(r)$ is the direct correlation function. For {\em all bounded potentials with a Gaussian core} we conjecture that
\be
h(r)\approx 0\,\,\,\,{\rm for}\,\,\,\,\rho>\rho_0(T)\,,
\label{eq3}
\ee
$\rho_0(T)$ being a characteristic density expected to be higher for lower $T$. Under this assumption, the HNC $c(r)$ reads, for $\rho>\rho_0$,
\be
c(r)=h(r)-\ln[1+h(r)]-\beta u(r)\approx -\beta u(r)\,,
\label{eq4}
\ee
that is, the HNC closure reduces to the random-phase approximation~\cite{Hansen}, and
\be
\rho k_BTK_T^{({\rm HNC})}\approx\frac{1}{1+\rho\beta\widetilde{u}(0)}\,.
\label{eq5}
\ee
If $\widetilde{u}(0)>0$ (stable DGM potential) then $K_T^{({\rm HNC})}$ is positive for $\rho>\rho_0$, meaning that the BL cannot extend to infinite $\rho$. Conversely, if $\widetilde{u}(0)<0$ (unstable DGM potential) then the right-hand side of Eq.\,\ref{eq5} changes sign continuously at a certain $T$-dependent density, implying that the approximate $K_T$ diverges at the BL, which is identified by
\be
\rho\beta\widetilde{u}(0)=-1\,\,\,\,\,\,\Longrightarrow\,\,\,\,\,\,k_BT=|\widetilde{u}(0)|\rho\,.
\label{eq6}
\ee
Equation (\ref{eq6}) is checked below against the computed BL.

%
%     FIGURE 1
%
\begin{figure}[!b]
\begin{center}
\includegraphics[width=7.0cm]{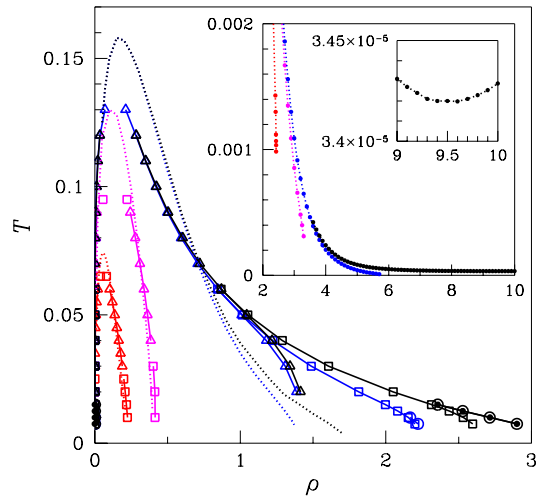}
\vspace{-10pt}
\caption{(Color online) DGM below $\eta_{\rm c}$: liquid-vapor coexistence densities from GEMC simulations performed for a number of $\eta$ values (0.020, red; 0.025, magenta; 0.0263, blue; and 0.02631, black). The various symbols correspond to different initial $N$ values in the two simulation boxes ($864+864$, triangles; $4000+108$, squares; $8192+128$, open dots; and $16000+128$, full dots). In each run, we generated a trajectory of $4\times 10^5$ MC cycles, of which only the second half was used to compute the averages. Also reported is the BL for the same $\eta$ values (dotted lines). Inset: BL for a few $\eta$ values close to $\eta_{\rm c}$ (0.026315, red; 0.0263157, magenta; 0.02631579, blue; and 0.02631580, black, further magnified in the top-right corner).}
\label{fig1}
\end{center}
\end{figure}

Far below $\eta_{\rm c}$, the DGM exhibits a liquid-vapor transition for very low $\rho$ and $T$~\cite{Prestipino1}, with a bell-shaped coexistence line whose height and width increase with $\eta$~\cite{Speranza}. Getting closer to $\eta_{\rm c}$, the binodal line undergoes a remarkable modification for low $T$, where it widens substantially (Fig.\,1), and a similar trend is shown by the BL of HNC theory. Crystallization can altogether be ignored near $\eta_{\rm c}$, since near the density $\rho_{\rm l}$ of the coexisting liquid, the crystal would only be stable for extremely low $T$~\cite{Speranza,Prestipino3}.

However, performing GEMC simulations very near $\eta_{\rm c}$ is a highly laborious task, since finite-size effects are relevant for low $T$. We thus followed the evolution of the BL as $\eta$ approaches $\eta_{\rm c}$ closer and closer (see Fig.\,1 inset). Apparently, $\rho_{\rm l}$ increases without bounds in the combined $\eta\rightarrow\eta_{\rm c}^-$ and $T\rightarrow 0$ limits; by contrast, for $\eta$ just above $\eta_{\rm c}$, the BL attains a low and flat minimum, after which it increases with density. Such findings are premonitory signs of unusual behavior beyond the stability threshold.

%
%     FIGURE 2
%
\begin{figure}
\begin{center}
\includegraphics[width=7.0cm]{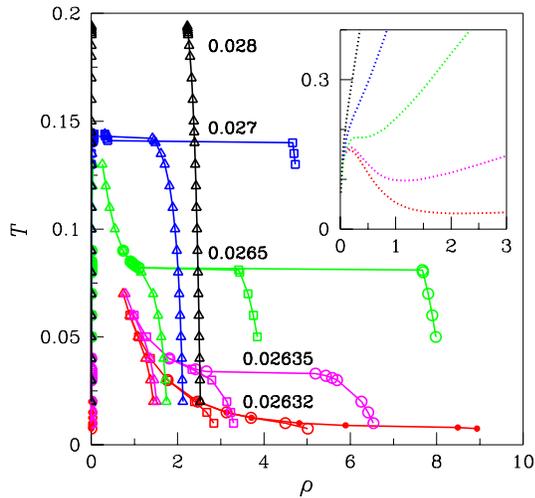}
\vspace{-10pt}
\caption{(Color online) DGM above $\eta_{\rm c}$: liquid-vapor coexistence densities from GEMC simulations performed for a number of $\eta$ values (0.02632, red; 0.02635, magenta; 0.0265, green; 0.027, blue; and 0.028, black --- same symbols as in Fig.\,1 caption). For $\eta=0.028$ no coexistence occurred above $T=0.194$. Inset: BL for a few $\eta$ values above $\eta_{\rm c}$ (from bottom to top: 0.02635, red; 0.0265, magenta; 0.027, green; 0.03, blue; and 0.05, black).}
\label{fig2}
\end{center}
\end{figure}

For $\eta>\eta_{\rm c}$, the region where the homogeneous fluid is unstable according to the HNC equation extends up to infinite density (Fig.\,2 inset). Above the BL, the DGM fluid would be homogenous. However, since the HNC theory is thermodynamically inconsistent, and thus blind to the system instability, this fluid can at best be {\em metastable}. In a narrow interval above $\eta_{\rm c}$ ($\eta_{\rm c}<\eta<\bar{\eta}\simeq 0.0270$) the BL exhibits a local maximum for $\rho\approx 0.2$, followed by a minimum at a higher density. Asymptotically, the BL becomes an ascending straight line. The two BL extrema coalesce for $\eta=\bar{\eta}$. As $\eta$ is increased further, the BL increasingly resembles a straight line, even for low densities. To a very high precision, the asymptotic BL slope coincides with $|\widetilde{u}(0)|$ for all $\eta>\eta_{\rm c}$, as expected from the conjecture (\ref{eq3}).

The above scenario is confirmed by simulation (see Fig.\,2). We first checked the existence of a (metastable) coexistence between liquid and vapor in the range $\eta_{\rm c}<\eta<\bar{\eta}$. We found that, above a temperature $T^*$ which is higher for larger $\eta$, coexistence densities are insensitive to $N$. Below $T^*$, coexistence is still apparently established but $\rho_{\rm l}$ now grows linearly with $N$, signaling that the denser phase is actually a non-extensive blob. For every $\eta$ the transition between the two cases is very sharp. In HNC theory, the analog of $T^*$ is the temperature of the BL local minimum.

%
%     FIGURE 3
%
\begin{figure}
\begin{center}
\includegraphics[width=7.0cm]{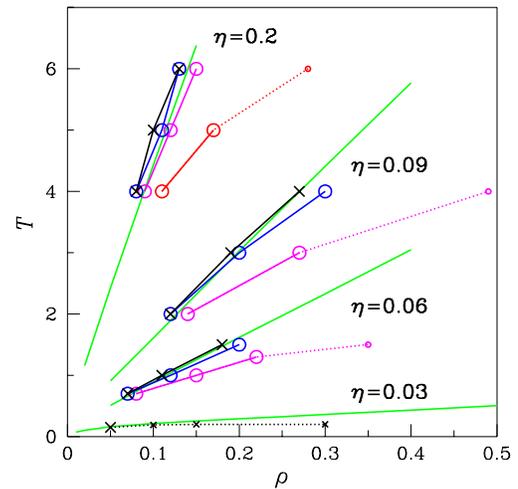}
\vspace{-10pt}
\caption{(Color online) DGM above $\eta_{\rm c}$: MC loci of the metastable-to-unstable crossover. Each color refers to a different $N$ (500, red; 1372, magenta; 4000, blue; and 8788, black crosses). In these simulations, $\rho$ was increased in steps of 0.01 at constant $T$ (for $\eta=0.03$, $T$ was reduced in steps of 0.01 at constant $\rho$). Any state point where collapse occurred abruptly was drawn as a large symbol, whereas a small symbol (dot or cross) marks the approximate location of a more diffuse (in density or in temperature) energy drop. For each $\eta$, the BL is plotted for comparison (green lines).}
\label{fig3}
\end{center}
\end{figure}

To better elucidate the nature of the crossover from metastable to unstable behavior, we carried out a few sequences of $NVT$ simulations as a function of $\rho$, for various choices of $\eta,T$, and $N$. We found that, at any given $T$, the average potential energy per particle, which is size-independent at low density, for a certain $N$-dependent density $\rho^*$ drops abruptly to a large negative value which, in absolute terms, scales as $N$~\cite{SM}. Hence, the total energy becomes of order $N^2$, which points to the formation of a non-extensive blob comprising all particles. The collapsed system appears as a spherical cluster, whose size is roughly independent of $N$, filling only a part of an otherwise empty simulation box~\cite{SM}. In Fig.\,3 we collected the $(\rho,T)$ points where system collapse first occurred. We see that, as $N$ grows, the crossover loci approach a definite limiting position which agrees well with the BL. This suggests that, for $\eta>\eta_{\rm c}$, the metastable-to-unstable crossover occurs at non-zero densities also in the thermodynamic limit, {\it i.e.}, there is a sharp line separating two distinct dynamic regimes: a ``long-lasting metastability'' regime where the time we should wait for collapse to manifest largely exceeds our typical simulation time (say, a few million MC cycles), and a regime of ``full-fledged instability'' where collapse occurs fast, well within the length of our runs.

A suggestive interpretation of the above findings, to be substantiated further on, is the following. The locus of $(\rho,T)$ points where, for $\eta>\eta_{\rm c}$, collapse is first observed is a sort of spinodal line (well approximated by the HNC BL) separating a metastable from an unstable region. Here, the phases competing for stability would be a {\em vanishing-density vapor} and a {\em huge-density liquid}. For $\rho<\rho^*(T)$, the homogeneous vapor is metastable. As $\rho$ overcomes $\rho^*$, the metastable vapor decays very quickly to a heterogeneous system made of a spherical huge-density-liquid cluster in a vanishing-density vapor.

In order to specify the position and width of the crossover region, we carried out a large number of MD simulations (for various densities, at fixed $\eta=0.2$ and $T=6$) where $N$ particles, originally distributed at random, are followed in the course of time until a sharp energy drop is detected~\cite{SM}. Away from the crossover region, MC averages do perfectly coincide with MD averages. In the crossover region differences might be observed due to the stochastic nature of the collapse event. 

%
%     FIGURE 4
%
\begin{figure}
\begin{center}
\includegraphics[width=7.0cm]{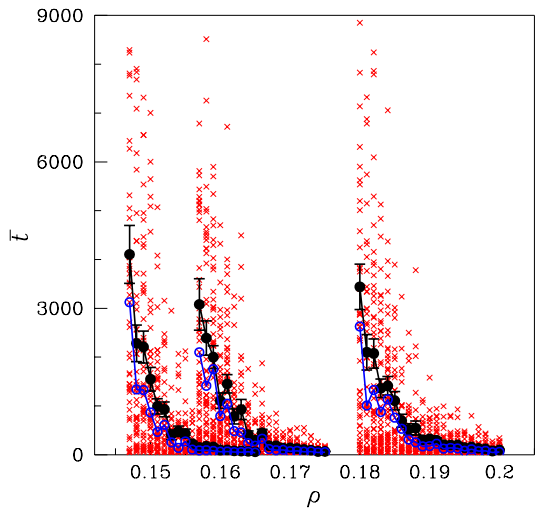}
%\vspace{-10pt}
\caption{(Color online) DGM for $\eta=0.2$ and $T=6$: collapse times (red crosses) for a large number of MD runs carried out in the metastable-to-unstable crossover region ($N=864$, right; $N=1372$, center; and $N=2048$, left). Some of the crosses fall outside the $t$ range shown. The black dots are arithmetic averages over each set of 31 instances (51, for every of the lowest five densities in each group). Also shown are the median times (open blue dots).}
\label{fig4}
\end{center}
\end{figure}

For each investigated density, we performed from 31 to 51 independent runs. In Fig.\,4 we show the mean waiting time for collapse as signaled by energy drop, $\bar{t}$, as a function of $\rho$, for three sizes ($N=864,1372,2048$). A distinct $\bar{t}$ increase is observed near the location of the collapse transition signaled by MC. For the lowest densities, the distribution $P(t)$ of waiting times is very dispersed around the mean and increasingly skewed to the right the smaller $\rho$ is. In principle $\bar{t}$ could blow up to infinity at a non-zero density or keep on increasing down to $\rho=0$. However, the trend of data in Fig.\,3 strongly points towards the former possibility~\cite{note}. The divergence of $\bar{t}$ at a non-zero density is consistent with $P(t)$ developing, below a certain density, a fat tail of the kind $t^{-\alpha}$, with $1<\alpha<2$ (in this case, $tP(t)$ is not integrable but $P(t)$ is normalizable). We argue that the existence of a sharp boundary separating the metastable from the unstable region finds a rationale in a different microscopic mechanism of collapse on the two sides of the locus: while on the low-density side aggregation would develop sequentially, through the repeated addition of particles until a critical nucleation cluster forms, on the other side of the boundary collapse is a sudden phenomenon with similarities to spinodal decomposition~\cite{SM}.

To delve more deeply into the nature of the collapsed state, we considered a system of particles interacting through the stabilized-DGM (SDGM) potential
\be
u_{\rm SDGM}(r)=\left\{
\begin{array}{rl}
\infty\,,
& r<\delta \\
u_{\rm DGM}(r)\,,  & r\ge\delta
\end{array}
\right.
\label{eq7}
\ee
with $\delta<\sigma$. This system behaves similarly to the DGM for low to intermediate densities, departing from it only for larger densities, where the inner core becomes effective. The SDGM system is {\em fully thermodynamically stable} (by Corollary B in Ref.~\cite{Fisher}). In Fig.\,5 we reported the BL of HNC theory for $\delta=0.1$ and $\eta=0.0265$. For small densities, this curve is indeed indistinguishable from the DGM BL. For higher densities the SDGM BL develops a broad maximum (not shown). Given the rough correspondence between the BL and a spinodal curve in thermodynamically stable systems, this maximum is the signature of a transition from a low-density liquid to a high-density liquid~\cite{Franzese}. For very high densities the stable phase is a crystal, and the onset of crystallization is expected to occur near the freezing point of hard spheres with diameter $\delta$ ($\rho_{\rm f}\approx 943$ for $\delta=0.1$).

%
%     FIGURE 5
%
\begin{figure}
\begin{center}
\includegraphics[width=7.0cm]{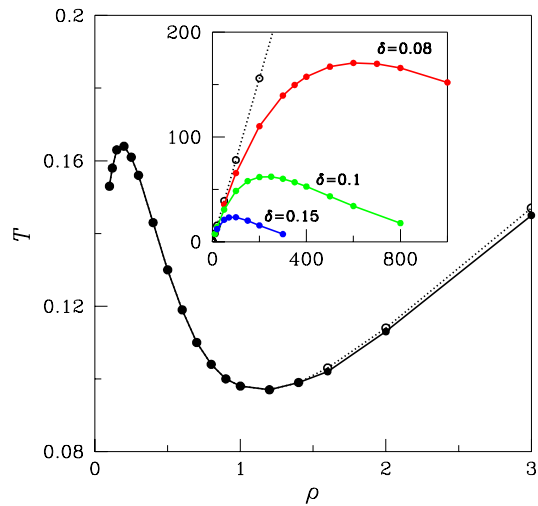}
\vspace{-10pt}
\caption{(Color online) SDGM for $\eta=0.0265$: BL for low densities ($\delta=0.1$, full dots; $\delta=0$, open dots). Inset: high-density behavior of the BL for $\eta=0.03$ and a number of $\delta$ values (0, black; 0.08, red; 0.1, green; and 0.15, blue).}
\label{fig5}
\end{center}
\end{figure}

We explored the high-density behavior of the BL as a function of $\delta$ for $\eta=0.03$ (see Fig.\,5 inset). As $\delta$ gets smaller, the BL maximum shifts to higher and higher $\rho$ and $T$. As a result, the ascending branch of the SDGM BL gets closer and closer to the DGM BL. Hence the locus which in the DGM separates the extensive from the non-extensive region can be viewed as the $\delta\rightarrow 0$ limit of the ascending branch of the SDGM fluid-fluid {\em spinodal} line. In turn, the metastable-to-unstable crossover may be interpreted as a sort of ``condensation'' where there is no repulsive core which prevents particles from crowding all together. In other words, the initial stages of collapse in the DGM resemble the onset of fluid-fluid separation in the SDGM, at least when the particle diameter is much smaller than the soft-core diameter. Also in regular condensation the formation of a spherical cluster is the first occurrence of liquid-vapor phase separation in the $NVT$-ensemble description~\cite{MacDowell,Abramo,Prestipino4}. The difference with the DGM cluster would only be in the radius, which for an ordinary liquid cluster in vapor scales as $N^{1/3}$ rather than staying finite.

The destabilization of a colloidal dispersion finds an interesting parallel in the SDGM phenomenology. Coagulation of a colloid can be triggered by suitably modifying the composition of the dispersion. In the schematization provided by the SDGM, the onset of aggregation would be nothing but the process where an increase of the attraction strength drives an originally homogeneous system inside the liquid-liquid spinodal region.

In conclusion, we investigated the phase diagram of a double-Gaussian fluid as a function of the attraction strength $\eta$. When the stability threshold $\eta_c$ is approached from the stable side, the liquid-vapor coexistence line undergoes a substantial widening for low temperatures, and the liquid density apparently diverges for $T\rightarrow 0$ right at $\eta_c$. Above $\eta_c$, a sharp line divides the thermodynamic plane in two regions characterized by radically different collapsing behaviors: very slow on one side of the line ({\it i.e.}, for low densities), extremely fast on the other side (high densities). The separatrix may be interpreted as the spinodal line of a transition between a vanishing-density vapor and a huge-density liquid.

\begin{acknowledgments}
We acknowledge useful discussions with Ezio Bruno, Dino Costa, Norberto Micali, and Valentina Villari.
\end{acknowledgments}

\end{document}